\documentclass[aps,pre,preprint,groupedaddress]{revtex4-1}

\usepackage{epstopdf}
\usepackage{empheq}
\usepackage{subfigure}
\usepackage{array}
\usepackage{graphicx}
\graphicspath{{figs/}}
\usepackage[dvips]{color}

\begin{document}


\title{Similarity transformation for equilibrium flows, including effects of blowing and suction}


\author{Xi Chen}
\email[]{xi.chen@ttu.edu}
\author{Fazle Hussain}
\affiliation{Department of Mechanical Engineering, Texas Tech University, Lubbock, TX,
79409, USA}


\date{\today}

\begin{abstract}
A similarity transformation for the mean velocity profiles is obtained in sink flow turbulent boundary layers (TBL), including effects of blowing and suction. It is based on symmetry analysis which transforms the governing partial differential equations (for mean mass and momentum) into an ordinary differential equation and yields a new result including an exact, linear relation between the mean normal ($V$) and streamwise ($U$) velocities. A characteristic length is further introduced which, under a first order expansion in wall blowing/suction velocity, leads to the similarity transformation for $U$. {This transformation is shown to be a group invariant under a generalized symmetry analysis} and maps different $U$ profiles under different blowing/suction conditions into a (universal) profile under no blowing/suction. Its inverse transformation enables predictions of all mean quantities in the mean mass and momentum equations - $U$, $V$ and the Reynolds shear stress - in good agreement with direct numerical simulation (DNS) data.
\end{abstract}

\pacs{}

\maketitle



\section{Introduction}
Equilibrium, denoted by the self-similarity (or self-preservation) of the mean profiles under proper normalization \citep{Townsend1976}, is one of the most fundamental concepts in turbulent boundary layers (TBLs). It is classified into two broad categories \citep{nickels2010}. One case involves an approximate equilibrium where the velocity and Reynolds
stresses are self-similar over most of the boundary layer and the other, an exact equilibrium where the self-similarity is observed over the entire layer thickness. The zero-pressure-gradient (ZPG) TBL, containing two independent (inner and outer) scales with the similarity properties expressed as the 'law of the wall' and 'defect law' \citep{marusic2010wall}, belongs to the first category. In contrast, the sink-flow boundary layer, a counterpart of the laminar Falkner-Skan boundary layers and a generic Jerrefy-Hamel flow constrained by two smooth plane surfaces, possesses many interesting properties \citep{jones2001,dixit2008}. This flow has an invariant velocity profile, a zero mean entrainment, radial mean streamlines, a constant Reynolds number ($Re$) and a constant friction coefficient along the stream, rendering it as the purest example of an exact equilibrium TBL \citep{spalart1986} and has triggered numerous studies on the scaling and flow structures \citep{coles1957,jones1972,jones2001,dixit2010}.

Whilst the laminar sink flow is one of the few known exact solutions of the Navier-Stokes equations \citep{Cantwellbook}, there is no known solution for the turbulent sink flow due to the Reynolds shear stress. This is analogous to the ZPG TBL, where various models are developed for the unclosed mean momentum equation. Notable works include an asymptotic logarithmic law for the mean velocity \citep{perry1994,jones2001, dixit2008}, the mixing length hypothesis \citep{patel1968}, etc. {However, to emphasize, except for the log-law, few are known for the sink flow TBLs. A crucial question concerns how the exact equilibrium state is produced \citep{spalart1986} and whether the state is robust under various boundary conditions. This is important because so far only the sink flow TBL is known to display the exact equilibrium state - first shown by \cite{Townsend1956} and \cite{Rotta1962}. To pursue more possible self-similarities in wall flows, a general theoretical framework is thus needed which is developed in this paper.}

Here, we use the Lie group symmetry analysis {\citep{Ibragimov1994,Cantwellbook}} to derive the self-similarity equation for boundary layers including effects of blowing and suction. It follows a recent work by \cite{Shesed1} with a notable difference, viz., it transforms the mean mass and mean momentum equations to an streamwise independent ordinary differential equation (ODE) and presents the necessary boundary conditions for the existence of the exact equilibrium state. It systematically unifies the Falkner-Skan laminar flows and the sink flow TBLs having different pressure gradients and blowing/suction, thus covering a wide class of equilibrium flows and perhaps fostering a more general study in future. {Note that recently two dimensional stagnation point flows have been discussed by Kolomenskiy \& Moffatt \cite{KM2012}, where a class of similarity solutions (exact equilibrium states) are obtained by (numerically) solving an ODE transformed from the two-dimensional Navier-Stokes equation. In fact, as will be shown later, their ODE and ours become the same under specific conditions.} Moreover,
the current approach is more general than previous scaling analysis \citep{Townsend1976, dixit2008} where specific scales are proposed for specific flows. Also note that while only the laminar cases (Blasius and Falkner-Skan) have been studied through a symmetry analysis by \cite{Cantwellbook}, {as well as the smooth wall turbulent boundary layers by \cite{Oberlack2001},} we extend the symmetry analysis to include the Reynolds shear stress and blowing/suction.

{More importantly, we establish a similarity transformation for the mean velocity profile covering ranges of blowing/suction strengths. {This transformation is found to be a group invariant (in most of the flow region except for the buffer layer) through a generalized symmetry analysis} and maps different $U$ profiles under different blowing/suction conditions into a universal profile under no blowing/suction. The latter reversely enables calculations of all quantities in the mean mass and momentum equations - in good agreement with DNS data \citep{Pat2014}. The results indicate that the wall blowing/suction not only preserves the equilibrium condition, but also leads to a new similarity among different blowing/suction strengths.}

The paper is thus organized as follows. Section II is devoted to a symmetry analysis of the mean mass and streamwise mean momentum equations, resulting a generalized ODE for various flows mentioned above. The similarity transformation for $U$ is presented in section III; also included is a prediction of the mean velocities and Reynolds stresses. Section IV presents the conclusions and discussions.

\section{Symmetry transformation for the boundary layer equations}

\begin{figure}
\centering
\includegraphics[width = 7 cm]{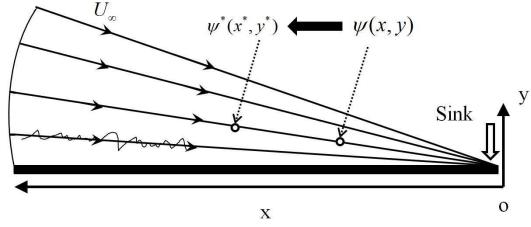} 
  \caption{Sketch of a sink flow turbulent boundary layer and the symmetry transformation from $\psi(x,y)$ to $\psi^*(x^*,y^*)$ following (\ref{eq:dilation}), i.e. from location $(x,y)$ to $(x^*,y^*)$.}
  \label{fig:sink}
\end{figure}

The incompressible, two-dimensional Navier-Stokes equations with the standard boundary layer approximation (NSBL) read
\begin{equation}\label{eq:mass}
\frac{{\partial U}}{{\partial x}} + \frac{{\partial V}}{{\partial y}} = 0
\end{equation}
\begin{equation}\label{eq:MME}
U\frac{{\partial U}}{{\partial x}} + V\frac{{\partial U}}{{\partial y}} = {U_\infty }\frac{{\partial {U_\infty }}}{{\partial x}} + \nu \frac{{{\partial ^2}U}}{{\partial {y^2}}} - \frac{{\partial R}}{{\partial y}}
\end{equation}
where $U$, $V$ indicate mean streamwise ($x$) and wall normal $y$ velocities; $R=\left<u'v'\right>$ is the Reynolds shear stress. Note that a zero $R$ indicates the laminar flow. The boundary conditions are $U(y = 0) = 0$, $U(y \rightarrow \infty) = U_\infty(x)$ and $V(y = 0) = {V_w(x)}$, where a zero $V_w$ indicates the non-penetrating wall and else for suction (a negative $V_w$) and blowing (a positive $V_w$) effects. Note that the origin location $(x,y)=(0,0)$ is set at the sink apex shown in figure \ref{fig:sink}, to explain the dilation transformation defined later.

Similar to the analysis of the laminar boundary layer flow, we introduce the streamfunction to eliminate the mass equation. The novelty is that, due to blowing/suction effects, the velocities are:
\begin{equation}\label{eq:psiA}
U = {\psi _y};\quad\quad V =  - {\psi _x} + V_w.
\end{equation}
One can check that (\ref{eq:psiA}) always satisfies (\ref{eq:mass}) as long as $\psi _{xy}=\psi _{yx}$, common to previous analysis \citep{Cantwellbook}. Therefore, (\ref{eq:MME}) written in stream function reads
\begin{equation}\label{eq:MMEpsi}
{\psi _y}{\psi _{xy}} - {\psi _x}{\psi _{yy}} +V_w{\psi _{yy}} = {U_\infty }{{\partial_x {U_\infty }}}  + \nu {\psi _{yyy}} - {R_y}
\end{equation}
where $V_w{\psi _{yy}}$ and ${R_y}=\partial_y R$ are additional compared to the Falkner-Skan equation. {Following the procedure in \cite{Cantwellbook}, here we search for the dilation symmetry permitted by (\ref{eq:MMEpsi}) (more discussion see appendix). Denoting
\begin{eqnarray}\label{eq:dilationser}
{x^*} &=& {e^{a_1} }x;\quad\quad\quad\quad {\rm{  }}{y^*} = {e^{{a_2}}}y;\quad\quad U_\infty ^* = {e^{{a_3}}}{U_\infty };\nonumber\\
V^*_w &=& {e^{{a_4}}} V_w; \quad\quad\quad  {\psi ^*} = {e^{{a_5}}}\psi;\quad\quad {\rm{  }}{R^*} = {e^{{a_6}}}R,
\end{eqnarray}
and substituting (\ref{eq:dilationser}) into (\ref{eq:MMEpsi}), we obtain
\begin{eqnarray}\label{eq:MMEpsi2}
e^{(2a_2+a_1-2a_5)}{\psi^\ast _{y^\ast}}{\psi^\ast _{x^\ast y^\ast}} - e^{(2a_2+a_1-2a_5)}{\psi^\ast _{x^\ast}}{\psi^\ast _{y^\ast y^\ast}} +e^{(2a_2-a_4-a_5)} V^\ast_w{\psi^\ast _{y^\ast y^\ast}} \nonumber\\= e^{(a_1-2a_3)}{U^\ast_\infty }{{\partial_{x^\ast} {U^\ast_\infty }}}  + e^{(3a_2-a_5)}\nu {\psi^\ast _{y^\ast y^\ast y^\ast}} - e^{(a_2-a_6)}{R^\ast_{y^\ast}}
\end{eqnarray}
The dilation symmetry requires:
\begin{equation}\label{eq:ai}
2a_2+a_1-2a_5=2a_2-a_4-a_5=a_1-2a_3=3a_2-a_5=a_2-a_6,
\end{equation}
where four of the six free coefficients, i.e. $a_i$ (i=1,2,\ldots,6), can be determined from (\ref{eq:ai}). Without losing generality, we denote the two free coefficients as $a_1=\varepsilon$ and $a_3=\beta \varepsilon$. Thus the other four are given as
\begin{equation}\label{eq:a2}
a_2=(1-\beta)\varepsilon/2;\quad a_4=(\beta-1)\varepsilon/2;\quad a_5=(1+\beta)\varepsilon/2;\quad a_6=(3\beta-1)\varepsilon/2.
\end{equation}
Substituting (\ref{eq:a2}) back into (\ref{eq:dilationser}) yields} the two-parameter ($\varepsilon$ and $\beta$) dilation symmetry group:
\begin{eqnarray}\label{eq:dilation}
{x^*} &=& {e^\varepsilon }x;\quad\quad\quad\quad {\rm{  }}{y^*} = {e^{(1 - \beta )\varepsilon /2}}y;\quad\quad U_\infty ^* = {e^{\beta \varepsilon }}{U_\infty };\nonumber\\
V^*_w &=& {e^{(\beta-1) \varepsilon/2 }} V_w; \quad  {\psi ^*} = {e^{(1 + \beta )\varepsilon /2}}\psi;\quad\quad {\rm{  }}{R^*} = {e^{(3\beta  - 1)\varepsilon /2}}R
\end{eqnarray}
which has a clear explanation, i.e. a mapping of a solution at the location $(x,y)$ to a series of solutions at locations $(x^*, y^*)$ in the sink flow (see figure \ref{fig:sink}). Note that translations for $x,y,\psi,R$ also keep (\ref{eq:MMEpsi}) invariant, which, however, break the invariance of wall conditions $\psi(x=0)=0$ and $R(y=0)=0$, hence not considered here.

An important fact is that the symmetry group (\ref{eq:dilation}) implies the necessary boundary conditions for the existence of the equilibrium state (i.e. the similarity solution) as follows. By integrating the characteristic equations of (\ref{eq:dilation}), i.e.
\begin{equation}\label{eq:cha}
\frac{dx}{x}=\frac{dy}{(1 - \beta )y/2}=\frac{{d}U_\infty}{\beta U_\infty}=\frac{{d}V_w}{(\beta -1)V_w/2}=\frac{{d}\psi}{(1 + \beta )\psi/2}=\frac{{d}R}{(3\beta -1)R/2},
\end{equation}
we obtain five independent dilation invariants:
\begin{eqnarray}\label{eq:invaraint}
{I_1} = y&/&{x^{(1 - \beta )/2}};\quad {I_2} =  {U_\infty }/{x^\beta };\nonumber\\
I_3 = V_w/{x^{(\beta-1 )/2}};\quad {I_4} &=& \psi /{x^{(1 + \beta )/2}};\quad \quad {I_5} = R/{x^{(3\beta  - 1)/2}},
\end{eqnarray}
which are explained in order. The first invariant $I_1$, in analogy to the similarity variable $\chi =y/\sqrt{\nu x/U_\infty}$ in the Blasius equation, describes the characteristic line of the dilation, i.e. $y=I_1{x^{(1 - \beta )/2}}$. Taking $\beta=-1$ we have $y=I_1 x$, corresponding to the radial mean streamline in figure \ref{fig:sink} \citep{dixit2008}. The second invariant $I_2$ indicates that the pressure gradient parameter $K_p\equiv\nu \partial_x U_\infty /U^2_\infty$, {widely used in literature (e.g., \cite{spalart1986,dixit2008}), must satisfy
\begin{equation}
K_p=\nu\beta/(I_2 x^{\beta+1})\propto x^{-1-\beta}\end{equation}
where $U_\infty =I_2 x^\beta$ is substituted}. When $\beta=-1$, we have $K_p=const.$, which is indeed the pressure gradient condition for the sink flow TBL (also note that $I_2=U_\infty x$ indicating the sink strength \citep{jones2001}). Moreover, $I_3$ requires a streamiwse dependent blowing/suction velocity $V_w\propto x^{(1-\beta)/2}$. Again, $\beta=-1$ leads to $V_w\propto U_\infty \propto 1/x$, which is exactly the blowing/suction setting in the DNS by \cite{Pat2014} for the sink flow TBL. Finally, $I_4$ and $I_5$ respectively indicate the invariants along the characteristic lines, composed of $\psi$ and $R$ (both are dependent variables) with $x$. Specifically for the sink flow, $\psi$ keeps invariant under dilation. Note that all of the group parameters are independent of viscosity (or $Re$). Such a $Re$-independent dilation invariance should be considered as a significant property of the sink flow TBL, because the symmetry could be physically identified in the flow field without changing the viscosity (by different fluids), as sketched in figure \ref{fig:sink}.

Furthermore, the PDE system (\ref{eq:mass})-(\ref{eq:MME}) is now transformed to an ODE under (\ref{eq:dilation}). Before we proceed,
it is natural to normalize above invariants to be dimensionless  using $\nu$ (viscosity) and $I_2$ (sink strength) \citep{Cantwellbook}, {which are:
\begin{eqnarray}\label{eq:I}
\alpha  = {I_1}\sqrt {-I_2/\nu }=y\sqrt{-U_\infty /(x\nu)}; \quad \gamma =I_3/\sqrt {-I_2\nu}=V_w/\sqrt{-U_\infty \nu /x}; \nonumber \\F= {I_4}/\sqrt{-I_2\nu }=\psi/\sqrt{-U_\infty\nu x}; \quad E = {I_5}/\sqrt {{-I_2^3}\nu }=R/\sqrt{-U^3_\infty \nu /x}
\end{eqnarray}}
(negative $I_2$ due to $U_\infty<0$ in figure \ref{fig:sink}). Substituting (\ref{eq:I}) and (\ref{eq:invaraint}) into (\ref{eq:MMEpsi}) we obtain:
\begin{equation}\label{eq:F}
{F_{\alpha \alpha \alpha }} + (1 + \beta )F{F_{\alpha \alpha }}/2 - \beta F_\alpha ^2 + \beta  + {E_\alpha } - \gamma {F_{\alpha \alpha }} = 0
\end{equation}
which describes a class of the self-preserving flows. {Here, $\gamma$ represents the dimensionless blowing/suction velocity}. For $\beta=0$ (ZPG), $E=0$ (laminar flow) and $\gamma=0$ (non-penetrating wall), (\ref{eq:F}) is the Blasius equation for laminar boundary layers. For nonzero $\beta$ with $E=\gamma=0$, (\ref{eq:F}) is the Falkner-Skan family of boundary layers, with an exact analytical solution for $\beta=-1$ (i.e. $F_\alpha=3 \tanh^2[\alpha+\tanh^{-1}\sqrt{2/3}]-2$) \citep{Cantwellbook}. {For $\beta=1$ and $E=\gamma=0$, (\ref{eq:F}) is the similarity equation obtained in \cite{KM2012} for stagnation point flows (Equation 2.2 with $\kappa=0$ indicating the steady flow condition).} Moreover, for $\beta=-1$ with nonzero $E$ and $\gamma$, (\ref{eq:MMEpsi}) becomes:
\begin{equation}\label{eq:STBLBS}
{F_{\alpha \alpha \alpha }} + F_\alpha ^2 - 1 + {E_\alpha } - \gamma {F_{\alpha \alpha }} = 0
\end{equation}
which is the self-preserving form of the sink flow TBL with blowing/sucntion effects. Note that (\ref{eq:STBLBS}) has been obtained by \cite{dixit2008} (for $\gamma=0$) and \cite{Pat2014} (for $\gamma\neq0$) by dimensional analysis (assuming a specific self-similarity). However, the symmetry analysis here is more straightforward (the advantage as emphasized in \citep{Cantwellbook}) resulting from the NSBL equation, and (\ref{eq:F}) is more general than (\ref{eq:STBLBS}) which indicates that there may exist other equilibrium flows with different values of $\beta$ and $\gamma$ (an open issue for future study).

\begin{figure}
\subfigure []{\includegraphics[width = 7 cm]{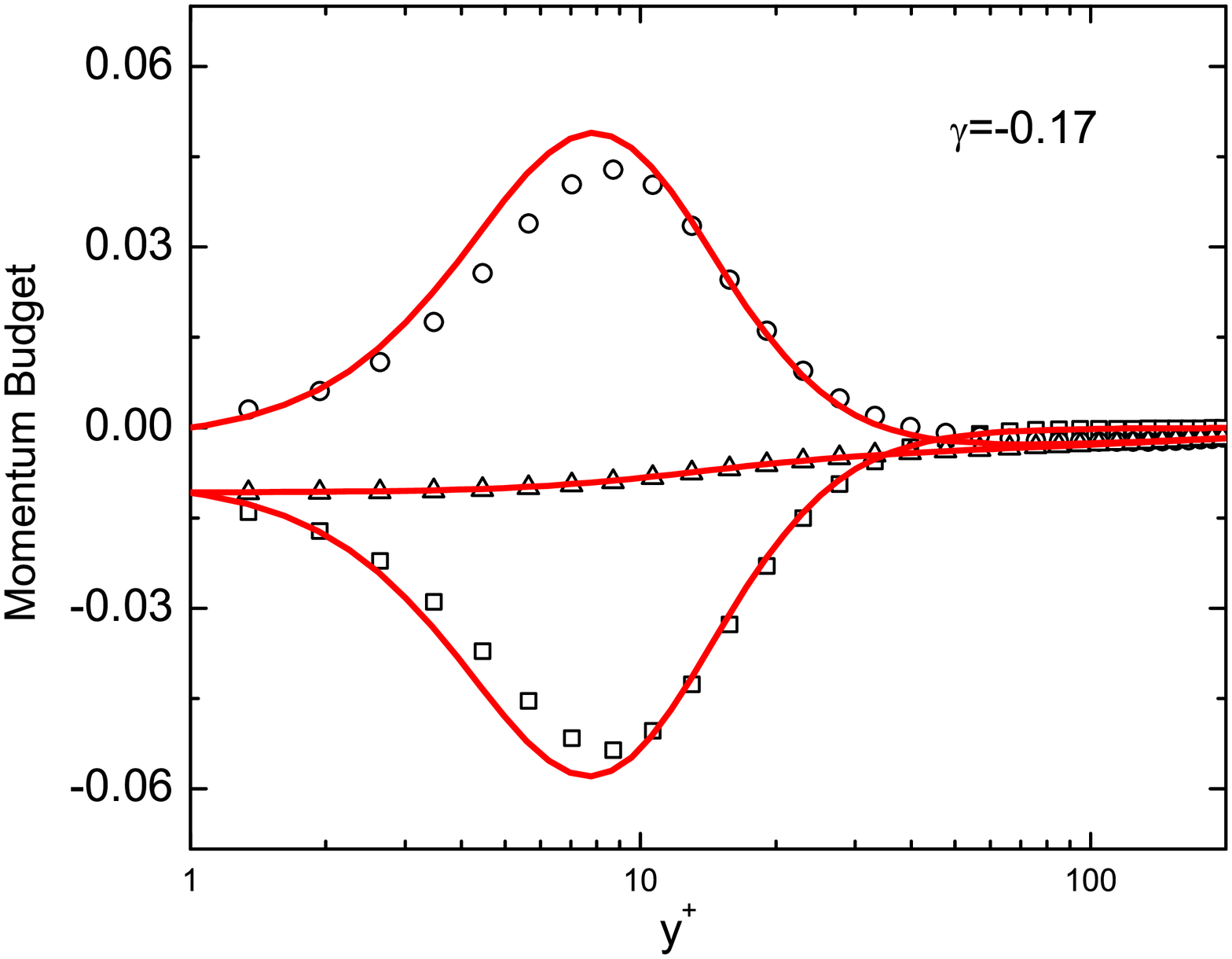}} 
\subfigure []{\includegraphics[width = 7 cm]{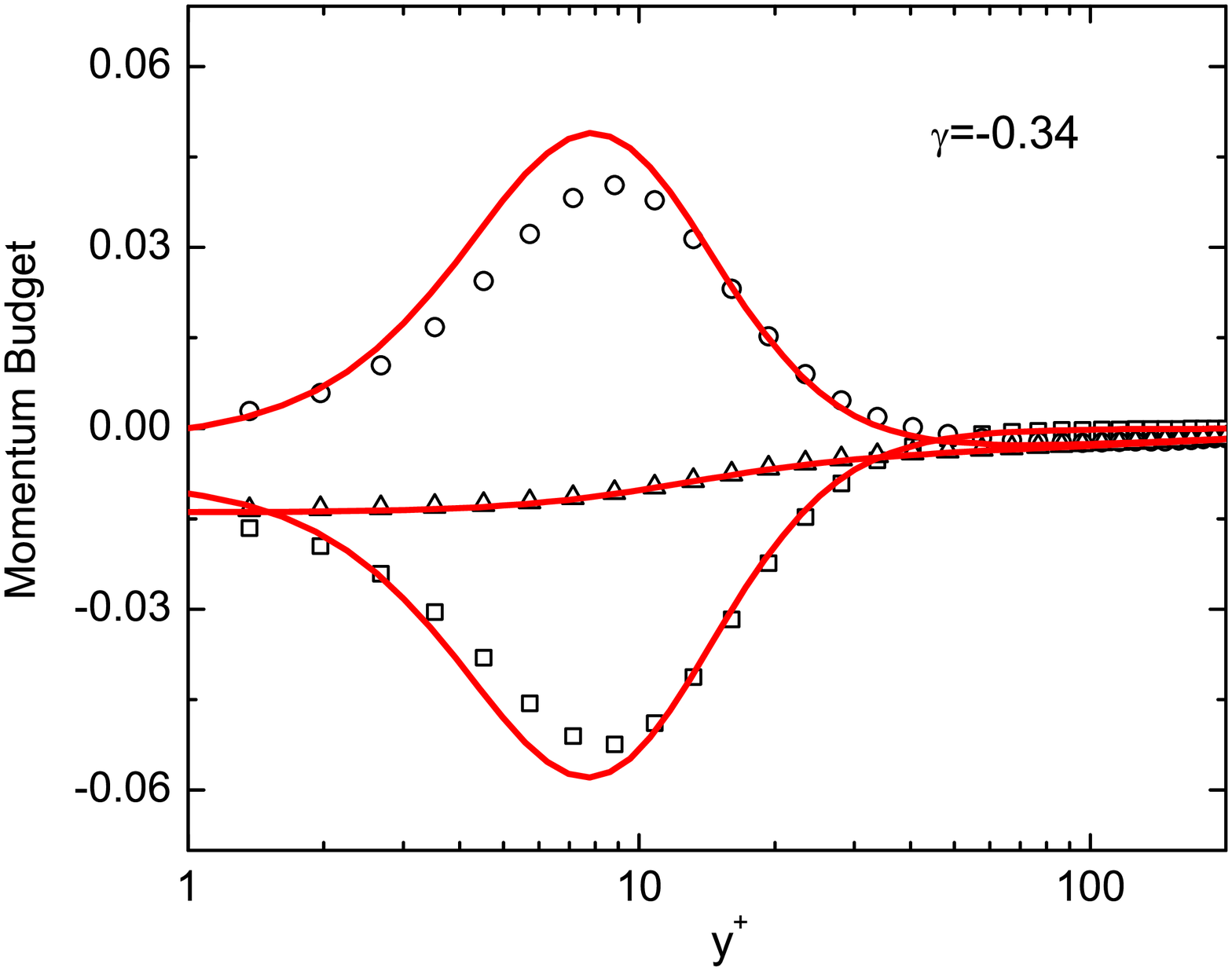}} 
\subfigure []{\includegraphics[width = 7 cm]{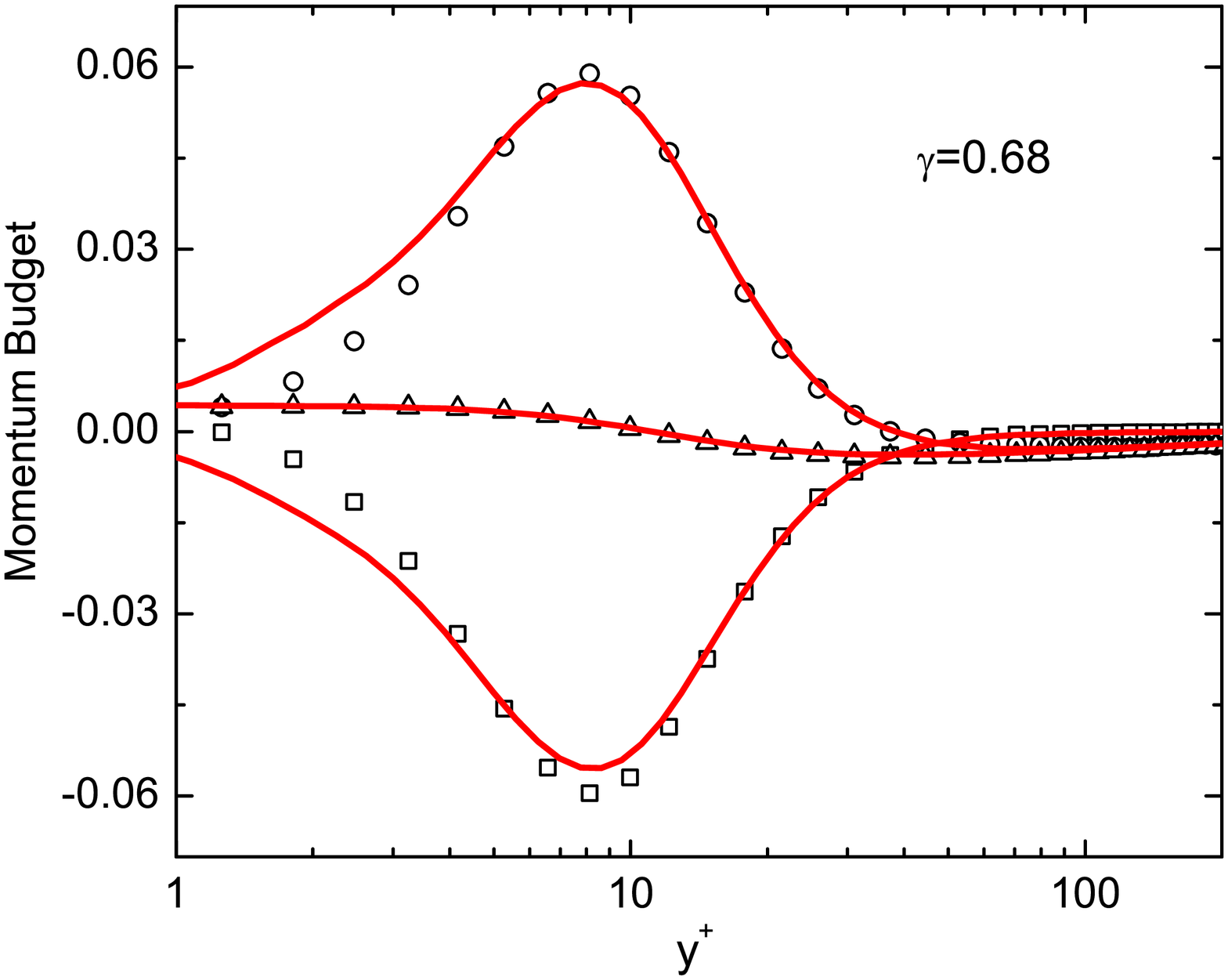}} 
\subfigure []{\includegraphics[width = 7 cm]{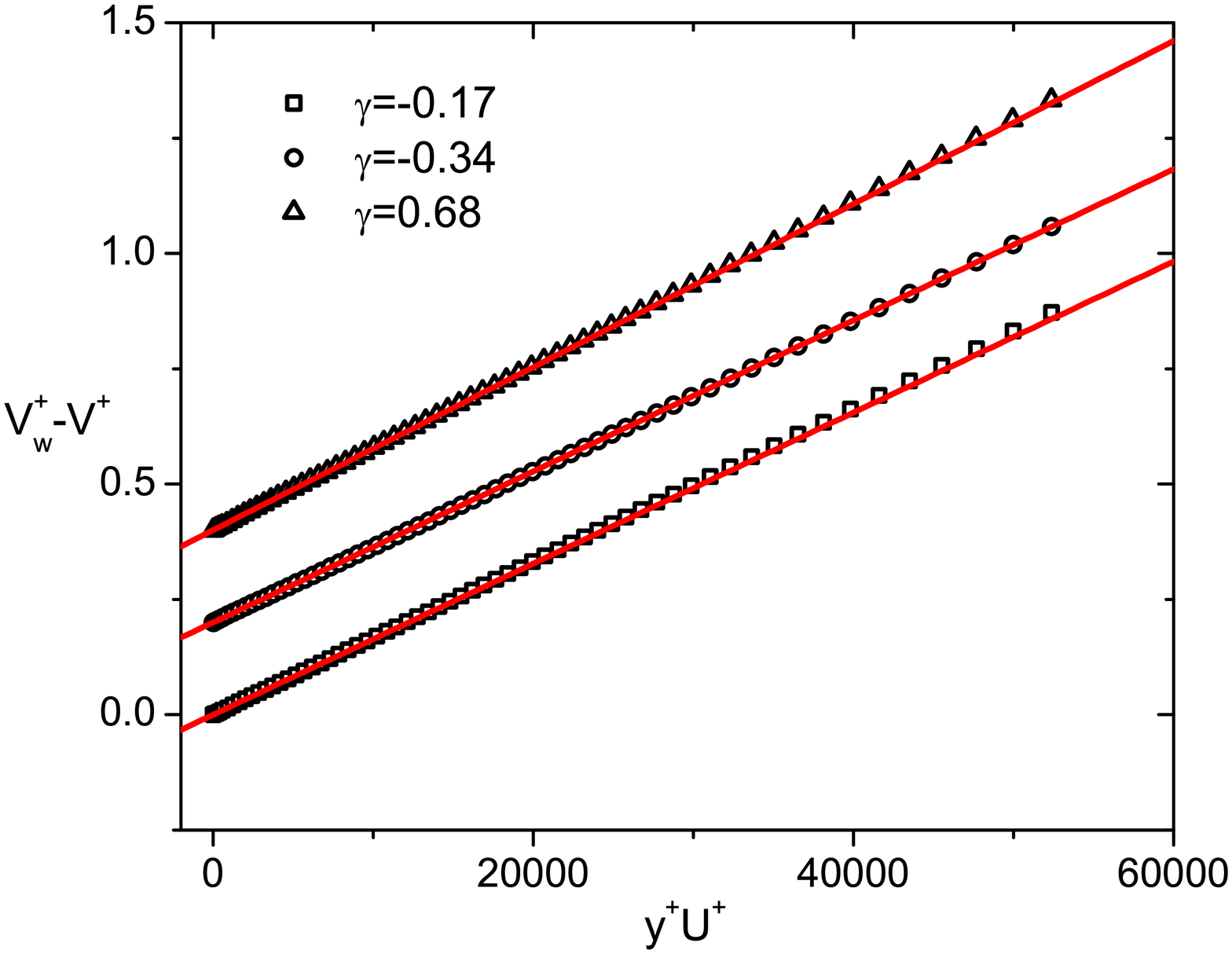}} 
  \caption{{(a)-(c) Budgets of (\ref{eq:MMEutau}). Symbols - DNS data \citep{Pat2014} for $\gamma=-0.17,-0.34,0.68$; lines - theoretical descriptions. Squares - $\partial^2 U^+/\partial y^{+2}$; circles - $\partial R^+/\partial y^+$; triangles - the right hand side of (\ref{eq:MMEutau})}. (d) Verification of the linear relation between $V^+$ and $U^+$ for different $\gamma$'s where lines denote (\ref{eq:UV}). Each profile has been vertically shifted by 0.2 for better display. }
  \label{fig:UV}
\end{figure}

Several interesting results can be deduced for $\beta=-1$. First, it is derived from the definition that the wall friction velocity $u_\tau\equiv\sqrt{-\nu{\partial_y U}\mid _{y=0}}=-U_\infty \sqrt{F_{\alpha\alpha}|_{\alpha=0} K_p}$ scales the same as the free stream velocity, i.e. $u_\tau\propto-U_\infty$ (since $K_p=\sqrt{-\nu x^{-1} U^{-1}_\infty}$ is a constant). Then, the similarity variable $\alpha$ (dimensionless invariant) in (\ref{eq:STBLBS}) actually scales the same as the viscous unit, i.e. $\alpha={y^ + } U^+_\infty \sqrt{K_p}\propto y^ +$, and (\ref{eq:STBLBS}) can be rewritten as:
\begin{equation}\label{eq:MMEutau}
\frac{{{\partial ^2}{U^ + }}}{{\partial {y^{ + 2}}}} + \frac{{\partial {R^ + }}}{{\partial {y^ + }}} = \gamma U_\infty ^ + \sqrt {{K_p}} \frac{{\partial {U^ + }}}{{\partial {y^ + }}} - {K_p}U_\infty^+ (U_\infty ^{ + 2} -{U^{ + 2}})
\end{equation}
where superscript $+$ denotes viscous normalization, i.e. ${y^ + } = y{u_\tau }/\nu$, ${U^ +_\infty } =  - U_\infty/{u_\tau }$, $U^+=-U/u_\tau$, ${R^ + } =  R/{u^2_\tau }$ and ${V^ + } = V/{u_\tau }$ (all normalized variables are positive). A validation of (\ref{eq:MMEutau}) is shown in figure \ref{fig:UV}, in good agreement with our theoretical descriptions (explained later). Note that superficially (\ref{eq:MMEutau}) shows no explicit $Re$ dependence, but in fact the latter is contained in the pressure gradient parameter $K_p$. In DNS data of \cite{Pat2014}, $K_P\approx7.71\times 10^{-7}$ is fixed while {the dimensionless blowing/suction strength} $\gamma=V_w/\sqrt{-U_\infty\nu/x}$ {(invariant along the stream)} varies within a typical range from $-0.34$ to $0.68$. This thus allows us to focus on the wall blowing/suction effects here, leaving the $K_p$ effect for future study (we hence omit the $K_p$ dependence below).

Moreover, {an exact relation between $U^+$ and $V^+$ following the streamfunction (\ref{eq:psiA}) is obtained}:
\begin{equation}\label{eq:UV}
{V^ + } = V_w^ + - {K_p}U_\infty ^ + {y^ + }{U^ + }
\end{equation}
Here $\psi=\nu F/\sqrt{K_p}$ and $\psi_x/u_\tau=K_p U^+_\infty y^+ U^+$ are substituted (note also $V_w^ +=V_w/u_\tau$). The comparison with DNS data \citep{Pat2014} is shown in figure \ref{fig:UV}. Note that the remarkable linear slope extends from wall to the entire flow region. The data agrees with the theoretical $K_p U^+_\infty$ in (\ref{eq:UV}) closely thus validating the above analysis.

\section{A similarity transformation for different $\gamma$'s}
We pursue the intriguing question, i.e. how the wall blowing/suction influences the mean velocity. To address this, let us recall a similar problem in compressible flows where the mean velocity is altered by the density variation. Through the well-known van Driest transformation, different mean velocities at different $Ma$'s are transformed into a universal profile at $Ma=0$ \citep{van1951,zhang2012}. Similarly, it is natural to make an analogy that mean velocities at different $\gamma's$ would be transformed to be the universal one at $\gamma=0$ - when wall blowing/suction effects are considered in a proper way. This is formally expressed as:
\begin{equation}\label{eq:transA}
U_*^ +(y^+, 0)  = \int {\phi {S^ +(y^+, \gamma) } dy+ }
\end{equation}
where $U_*^+$ is the mean velocity at $\gamma=0$, i.e. $U_*^+ =U^ +(y^+, 0)$; $\phi (y^+,\gamma)$ is the weighting function, and $S^+({y^ + }, \gamma)=\partial_{y^+} U^+$ is the mean shear obtained from mean velocity profiles $U^ +(y^+, \gamma)$ for blowing/suction conditions. Note that $\phi$ is a function of mean density in the van Driest transformation; here $\phi$ is unknown a priori, whose determination (as below) thus achieves a quantitative characterization of the intriguing blowing/suction effects. To emphasize, the existence of such a $\phi$ in (\ref{eq:transA}) is nontrivial, because it requires that the wall blowing/suction not only preserves the streamwise equilibrium condition, but also leads to a new similarity among different $\gamma$'s - never addressed before.

In fact, (\ref{eq:transA}) has a transparent physical meaning. To see this, let us differentiate (\ref{eq:transA}) with $y^+$ and obtain $\phi^{-1}= S^+/S^+_*$. The latter indicates that $\phi^{-1}$ is the relative variation of the mean shear $S^+({y^ + }, \gamma)$ divided by $S^+_*={S^ + }({y^ + }, 0)$ at the non-penetrating wall. This is very much like the case in the rough pipes \citep{jimenez2004,shenjp}, where the mean flux in rough pipes subtracted by the smooth wall flux (the so called Hamas function) is the right quantity to reveal the similarity induced by roughness elements. This also inspires us to seek the expression of the $\phi$ function.

{Note that the transformation (\ref{eq:transA}) implies a $\gamma$-independent quantity $\phi S^+$ ($=S^+_*$). It motivates us to study the generalized symmetry of (\ref{eq:MMEutau}) where group invariants independent of $\gamma$ can be calculated from first principle. To extend the symmetries of (\ref{eq:MMEutau}), we take the derivative of (\ref{eq:MMEutau}) with respect to $y^+$ to obtain the PDE system \begin{eqnarray}\label{eq:PDE}
{S^ + } &=& {\partial _{{y^ + }}}{U^ + }\nonumber;\\
\frac{{{\partial ^2}{S^ + }}}{{\partial {y^{ + 2}}}} + \frac{{{\partial ^2}{R^ + }}}{{\partial {y^{ + 2}}}} &=& \gamma U_\infty ^ + \sqrt {{K_p}} \frac{{\partial {S^ + }}}{{\partial {y^ + }}} + 2{K_p}U_\infty ^ + {U^ + }{S^ + }.
\end{eqnarray}
Here, $K_p$ is a constant (pressure gradient); $U^+_\infty$ depends on $\gamma$; and $U^+, S^+$ and $R^+$ depend on $\gamma$ and $y^+$. Then, using Maple, the infinitesimals for the symmetry transformation of (\ref{eq:PDE}) are calculated:
\begin{eqnarray}\label{eq:GS}
\begin{array}{l}
{{\xi '}_\gamma } = {F_1},\\
{{\xi '}_{{y^ + }}} = ({F_2}/2){y^{ + 2}} + {F_3}{y^ + } + {F_4},\\
{{\eta '}_{{U^ + }}} =  - \gamma {F_2}/(2\sqrt {{K_p}} ) + ({F_3} + {F_5}){U^ + },\\
{{\eta '}_{{S^ + }}} = {S^ + }({F_5} - {F_2}{y^ + }),\\
{{\eta '}_{{R^ + }}} = {U^ + }U_\infty ^ + \sqrt {{K_p}} {F_1} + (3{S^ + }/2 + {R^ + }/2 - \gamma {U^ + }U_\infty ^ + \sqrt {{K_p}} ){y^ + }{F_2}\\
{\rm{        }} \quad\quad- 2\gamma {U^ + }U_\infty ^ + \sqrt {{K_p}} {F_3} + ({R^ + } - \gamma {U^ + }U_\infty ^ + \sqrt {{K_p}} ){F_5} + ({R^ + } + {S^ + }){F_6} + {F_7}{y^ + } + {F_8},\\
{{\eta '}_{U_\infty ^ + }} =  - ({y^ + }{F_2}/2 + 3{F_3} + {F_5} - {F_6})U_\infty ^ +
\end{array}
\end{eqnarray}
where $F_i$ ($i=1,...8$) are arbitrary functions of $\gamma$ and $U^+_\infty$.
The corresponding characteristic equations for group invariants are:
\begin{equation}\label{eq:CR}
\frac{{d\gamma }}{{{F_1}}} = \frac{{d{S^ + }}}{{({F_5} - {F_2}{y^ + }){S^ + }}} = \frac{{d{y^ + }}}{{({F_2}/2){y^{ + 2}} + {F_3}{y^ + } + {F_4}}}  =  \ldots
\end{equation}
Since our goal is a $\gamma$-independent $\phi S^+$, we focus on the invariant composed of $\gamma$ and $S^+$ by integrating the first equation in (\ref{eq:CR}), i.e.
\begin{eqnarray}\label{eq:IS2}
{I_S} &=& \ln ({S^ + }) - \int {({F_5}/{F_1})d} \gamma  + {y^ + }\int {({F_2}/{F_1})d} \gamma.
\end{eqnarray}
Therefore, $I_s$ and any functions of $I_s$ are also group invariants independent of $\gamma$. While this gives a general expression of a $\gamma$-independent quantity, we need to further identify the explicit expression of $\phi S^+$. As shown later below, $\phi S^+$ is indeed a function of $I_s$ and hence also a group invariant in most of the flow region (where $S^+\ll1$ or $S^+\approx1$). This is important because it supports that $S^+_*=\phi S^+$ is indeed a $\gamma$-independent quantity based on the first principle (i.e. the symmetry of (\ref{eq:PDE})). }

Below we start to derive an analytical $\phi$ once the mean velocity profile $U^+(y^+,\gamma)$ is known. At first, integrating (\ref{eq:MMEutau}) from $0$ to $y^+$ yields,
\begin{equation}\label{eq:SWT}
S^+ + {M^ + }= 1
\end{equation}
where $M^ +$ is the sum of the  shear stress (RS), the pressure gradient effect (PG) and the mean vertical convection (VC), i.e.
\begin{equation}\label{What}
{M^ + }({y^ + },\gamma) = \underbrace {{R^ + }}_{{\rm{RS}}} + \underbrace {{K_p}{y^ + }U_\infty ^{ + 3}}_{PG}\underbrace { - \gamma \sqrt {{K_p}} U_\infty ^ + {U^ + } - {K_p}U_\infty ^ + \int_0^{{y^ + }} {{U^{ + 2}}} dy'}_{VC}
\end{equation}
By dimensional analysis, a characteristic length function is introduced inspired by the order function concept in \cite{shenjp}:
\begin{equation}\label{eq:ell}
\ell^+(y^+,\gamma)=\sqrt{{M}^+}/{S^+}=\sqrt{1-S^+}/{S^+}
\end{equation}
therefore
\begin{equation}\label{eq:ratio}
\frac{l^ + }{l^ + _*}=\frac{\sqrt{1-S^+}/S^+}{\sqrt{1-S^+_*}/S^+_*}=\frac{\phi\sqrt{1-S^+}}{ \sqrt{1-\phi S^+}}
\end{equation}
where ${l^ + _*} = {l^ + }({y^ + },0)$. Then, (\ref{eq:ratio}) leads to an important expression for $\phi$ in terms of $S^+$ and $\ell^+/\ell^+_*$, i.e.
\begin{eqnarray}\label{eq:phiA}
\phi = {{{2\xi }}}/[{1 +  \sqrt {1 + 4{}\xi (\xi  - 1)/({\ell^+/\ell^+_*})^2} }]
\end{eqnarray}
where $\xi=1/S^+$ ($=1/\partial_{y^+} U^+$). Now, the key is to estimate $\ell^+/\ell^+_*$ as below. Considering that the moderate blowing/suction effect is indicated by a small parameter $|\gamma|<1$ (validated by all the data here), an expansion of $\ell^+(y^+,\gamma)$ in $\gamma$ is thus:
\begin{equation}
{l^ + }({y^ + },\gamma)={l^ + _*}(1 + \eta \gamma+\eta' \gamma^2+ h.o.t.)
\end{equation}
where coefficients $\eta=\partial_\gamma (\ell^+/\ell^+_*)|_{\gamma=0}$ and $\eta'=\frac{1}{2}\partial_\gamma[\partial_\gamma (\ell^+/\ell^+_*)]|_{\gamma=0}$ are generally functions of $y^+$. For simplicity, the expansions are truncated at the first order, i.e. ${l^ + }\approx{l^ + _*} (1 + \eta \gamma)$, which, after a substitution into (\ref{eq:phiA}), yields
\begin{eqnarray}\label{eq:phiB}
\phi \approx{{2\xi }} /[{1 +  \sqrt {1 + 4{}\xi (\xi  - 1)/{(1 + \eta \gamma)}^2} }]
\end{eqnarray}
and hence
\begin{eqnarray}\label{eq:Snew}
S^+_*=\phi S^+ \approx{{2}} /[{1 +  \sqrt {1 + 4{}\xi (\xi  - 1)/{(1 + \eta \gamma)}^2} }].
\end{eqnarray}

\begin{figure}
\subfigure []{\includegraphics[width = 7 cm]{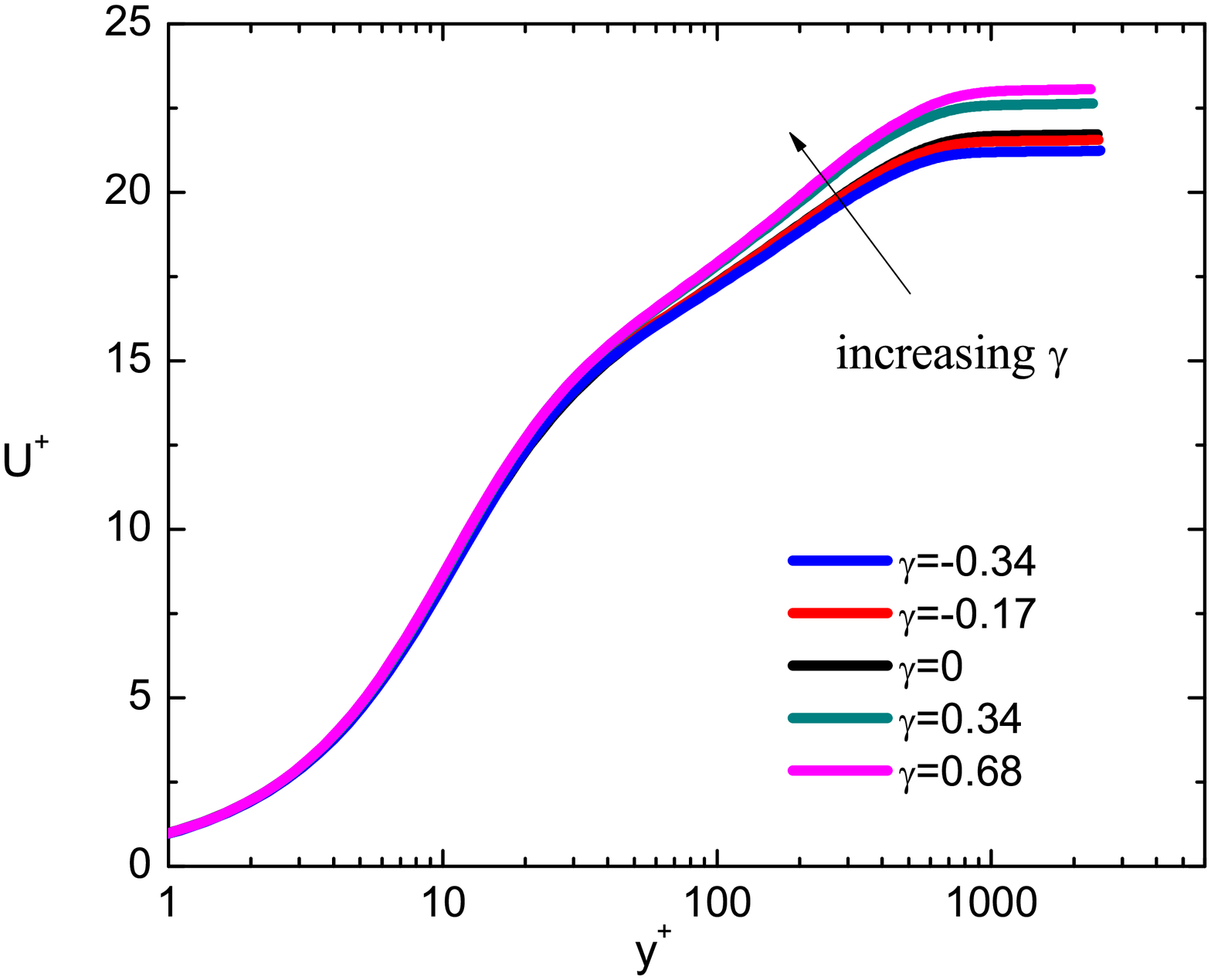}} 
\subfigure []{\includegraphics[width = 7 cm]{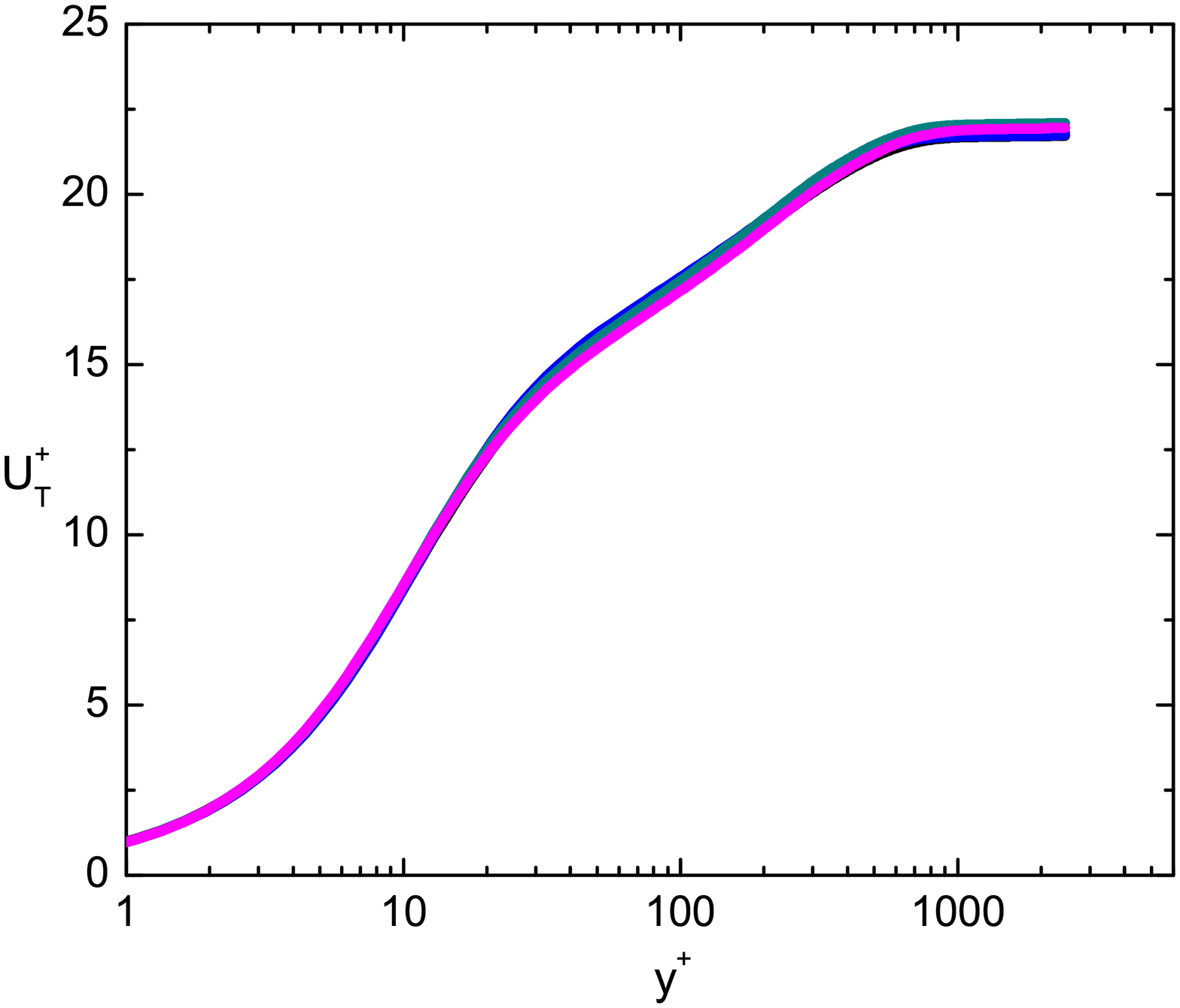}} 
\subfigure []{\includegraphics[width = 7 cm]{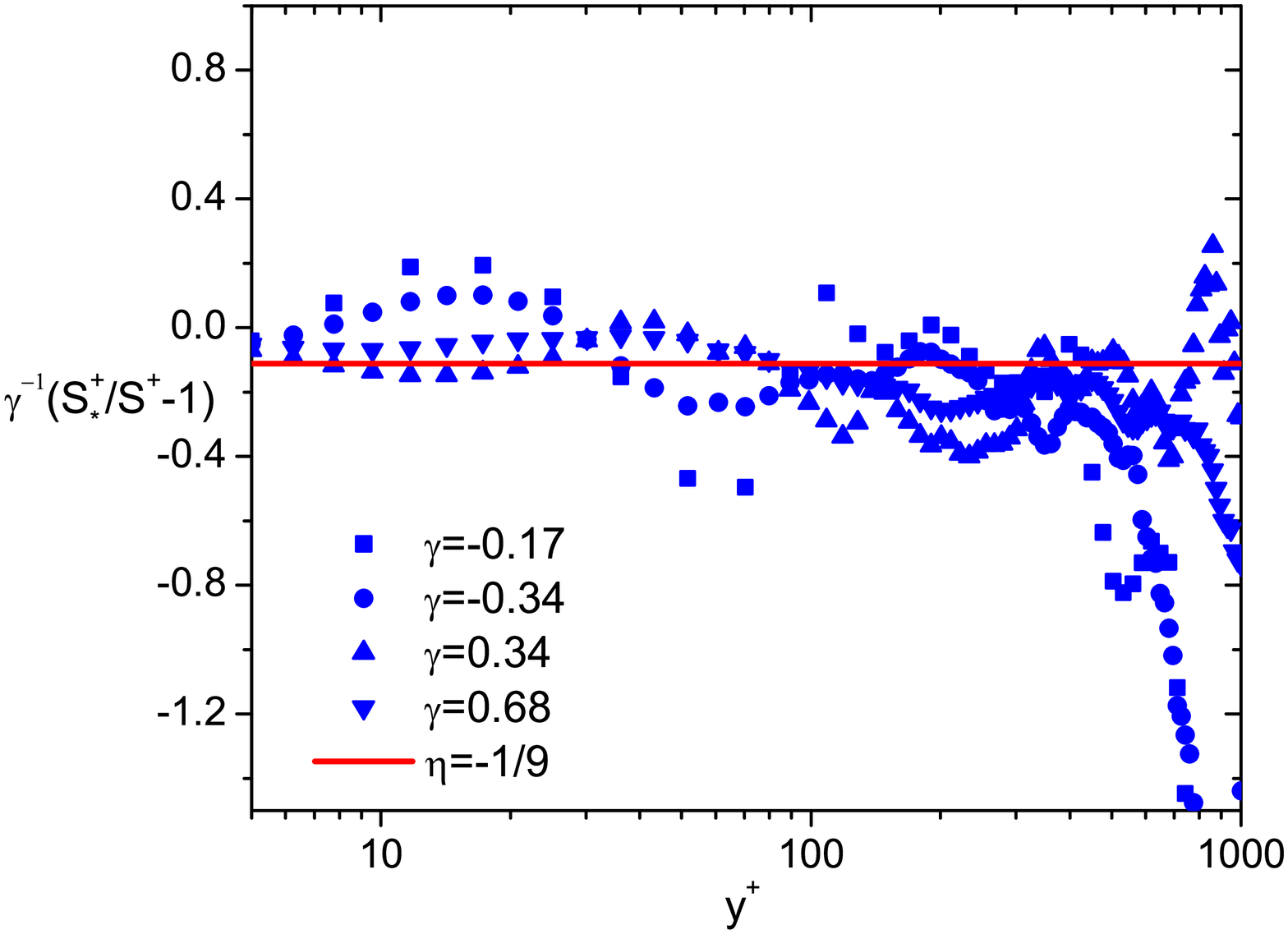}} 
\subfigure []{\includegraphics[width = 7 cm]{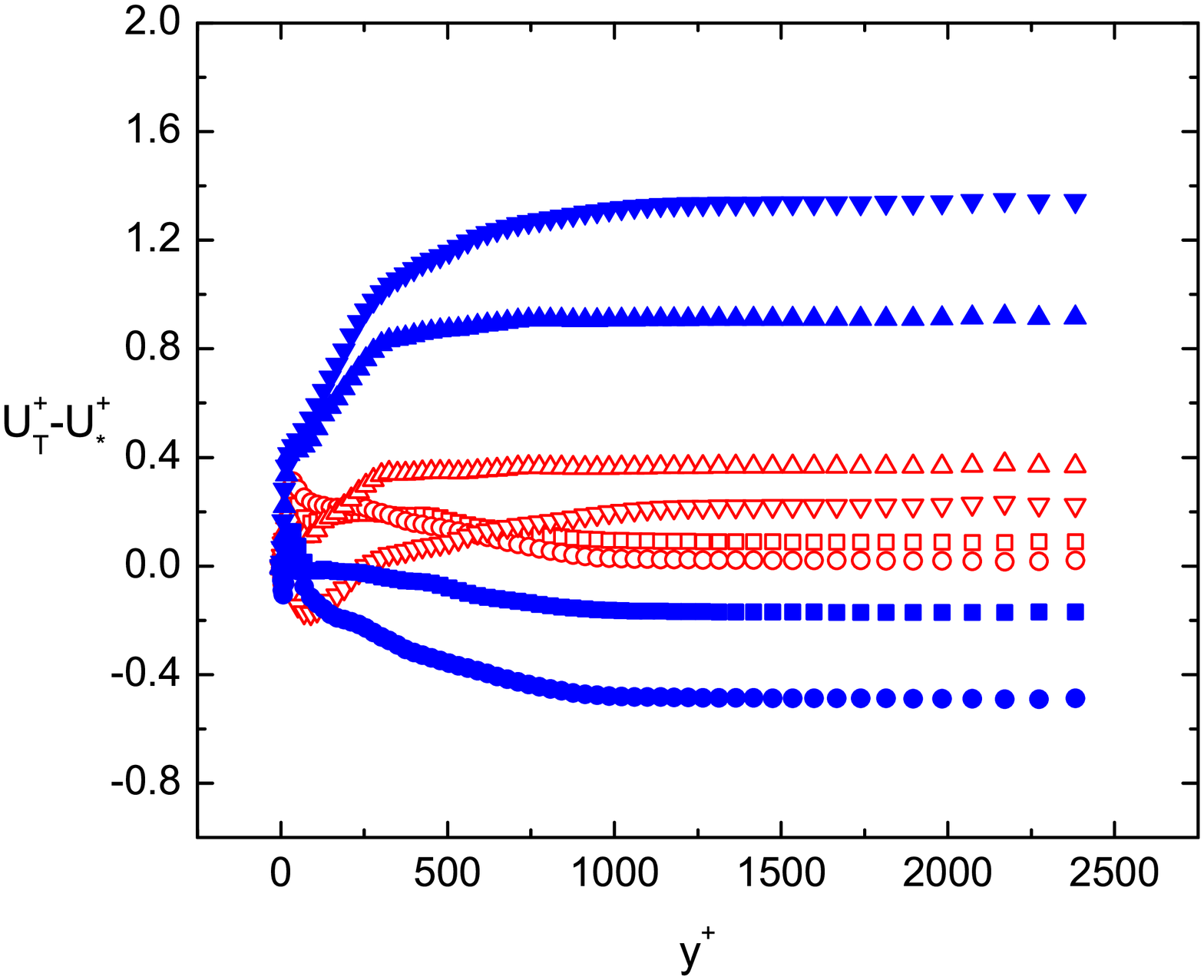}}      
\begin{center}{\subfigure []{\includegraphics[width = 7 cm]{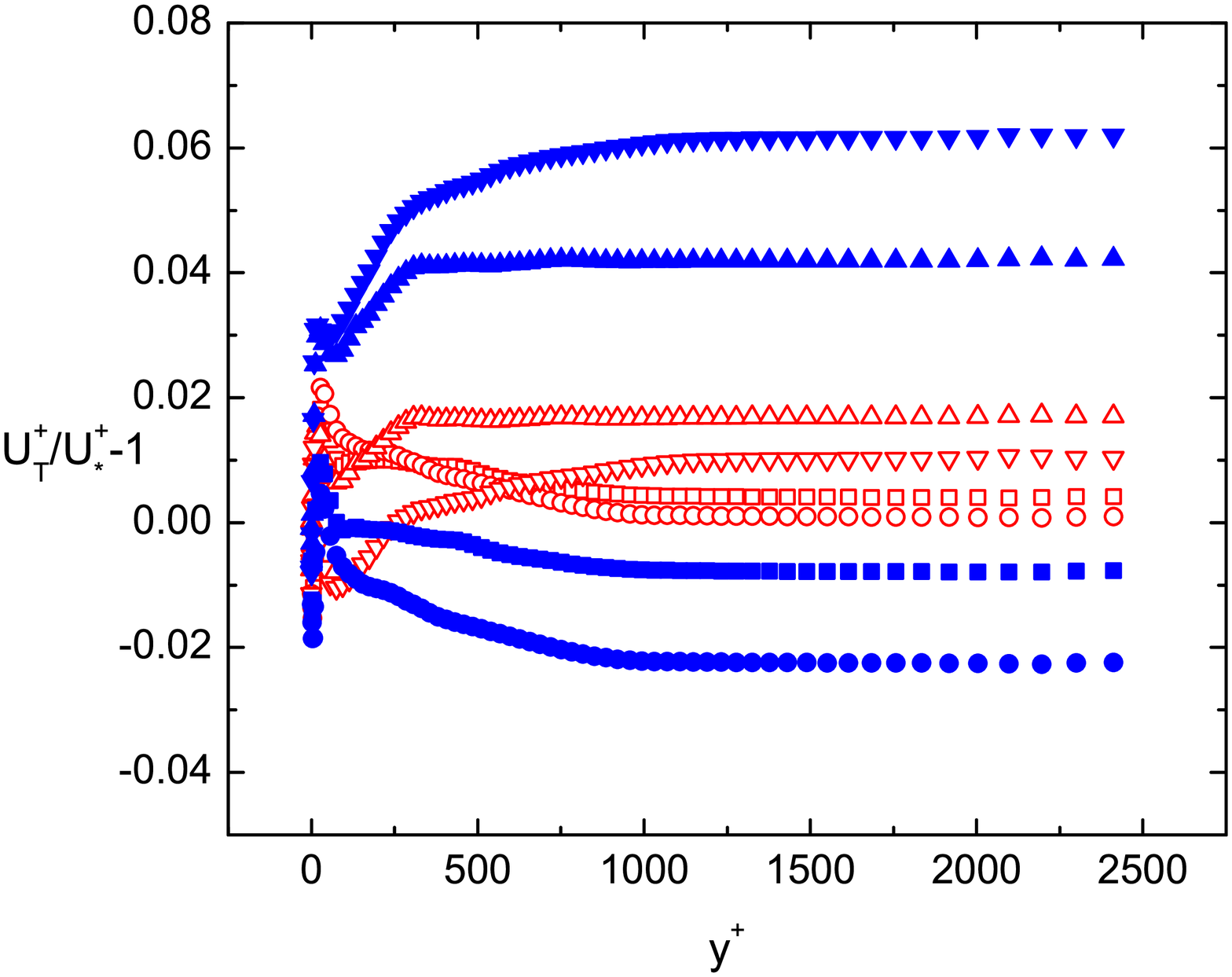}}}\end{center}
  \caption{A comparison of mean velocity profiles for different $\gamma$'s before (a) and after (b) transformations. (c) Measurement of $\eta=-1/9$ (line) by plotting $\gamma^{-1}(S^+_*/ S^+-1)$ using DNS data. Note that scatters towards freestream are due to $S^+_*$ and $S^+$ approaching zero (hence $S^+_*/ S^+$ is very sensitive to data). A comparison of the departures before and after transformation, i.e. $U^+-U^+_*$ (solids) versus $U^+_T-U^+_*$ (opens) is shown in (d); {and $U^+/U^+_*-1$ (solids) versus $U^+_T/U^+_*-1$ (opens) shown in (e).}}
  \label{fig:trans}
\end{figure}

{The relation between (\ref{eq:Snew}) and the symmetries of (\ref{eq:PDE}) is further discussed here. Note that for $\xi=1/S^+\gg1$ or $S^+\ll1$ (data showing $S^+<0.1$ beyond the buffer layer thickness $y^+\approx30$), (\ref{eq:Snew}) approximates to
\begin{equation}\label{eq:S12}
S^+_*=\phi S^+\approx{{2}} /[{1 +  \sqrt {1 + 4{}\xi^2/{(1 + \eta \gamma)}^2} }].
\end{equation}
Importantly, $\phi S^+$ in (\ref{eq:S12}) is indeed a function of the group invariant $I_S$, i.e.
\begin{equation}\label{eq:Isnew}
\phi S^+\approx{{2}} /[{1 +  \sqrt {1 + 4{}\xi^2/{(1 + \eta \gamma)}^2} }]={{2}} /[{1 +  \sqrt {1 + 4 /(e^ {I_s})^2}}].
\end{equation}
Here, after substituting the following specific conditions in (\ref{eq:IS2}), i.e.
\begin{equation}\label{eq:F5F2}
{F_5} =  - \eta {F_1}/(1 + \gamma \eta );\quad\quad {F_2} = 0,
\end{equation}
the invariant ${I_S}$ is given as:
\begin{equation}
{I_S} = \ln ({S^ + }) + \ln (1 + \gamma \eta ).
\end{equation}
Therefore, (\ref{eq:Isnew}) tells us that $\phi S^+$ in (\ref{eq:S12}) is also a group invariant hence independent of $\gamma$ and equals $S^+_*$. Note that equations (\ref{eq:GS}) with (\ref{eq:F5F2}) are called generalized symmetries, because they are permitted by (\ref{eq:PDE}), but not by (\ref{eq:MMEutau}) due to the cubic term $U^{+3}_\infty$ in the latter. Furthermore, the above analysis also applies to $\xi\approx S^+\approx1$ (viscous sublayer), where (\ref{eq:Snew}) approximates to $S^+_*\approx S^+\approx1$. The latter is of course a group invariant with $F_5=F_2=0$ in (\ref{eq:GS}); this result trivially follows from the linear expansion $U^+_*\approx U^+\approx  y^+$ at the wall (see figures \ref{fig:trans}a \& b).}

The value of $\eta$ is measured as below. Note that from (\ref{eq:phiB}), $\phi\approx 1+\eta \gamma$ for $\xi\gg 1$, indicating $\eta\approx \gamma^{-1}(\phi-1)$ for $y^+\gg1$. Therefore, we use DNS to measure the value of $\eta$ by plotting $\gamma^{-1}(S^+_*/ S^+-1)$ (see figure \ref{fig:trans}c), which is suggested as
\begin{equation}\label{eq:eta}
\eta=-1/9
\end{equation}
Here the negative sign indicates that $\ell^+$ decreases with increasing $\gamma$, which is physical. This is because a larger $\gamma$ indicates more mean vertical convection to be compensated by pressure force in (\ref{What}), hence a relatively smaller $M^+$ and a smaller $\ell^+$. Also note that the value $1/9$ is empirical whose $K_p$ dependence is an interesting topic for future study.

Therefore, we obtain the final transformation by substituting (\ref{eq:phiB}) with (\ref{eq:eta}) into (\ref{eq:transA}). Figure \ref{fig:trans}a shows notable departures of $U^+$ profiles from each other before the transformation, and the departures increase apparently with increasing wall distance. In contrast, figure \ref{fig:trans}b shows the transformed velocities ($U^+_T$) according to (\ref{eq:transA}), which remarkably collapse onto the universal one $U^+_*$ for the entire flow region. To display the quality of the collapse, figure \ref{fig:trans}d compares $U^+_T-U^+_*$ (opens) with $U^+-U^+_*$ (solids). While the maximum difference among two velocity profiles before transformation is between $\gamma=-0.3417$ and $\gamma=0.6834$, i.e. $\Delta U_{max}=U^+(\infty, 0.6834)-U^+(\infty,-0.3417)\approx 1.8$; differences after transformation are mostly bounded within 0.4. {We further plot $U^+_T/U^+_*-1$ in figure \ref{fig:trans}e, which are all bounded within 2\% after transformation.} This is satisfactory since we only use a the linear expansion for $\ell^+/\ell^+_*$ (i.e. a constant $\eta$).

\begin{figure}
\subfigure []{\includegraphics[width = 7 cm]{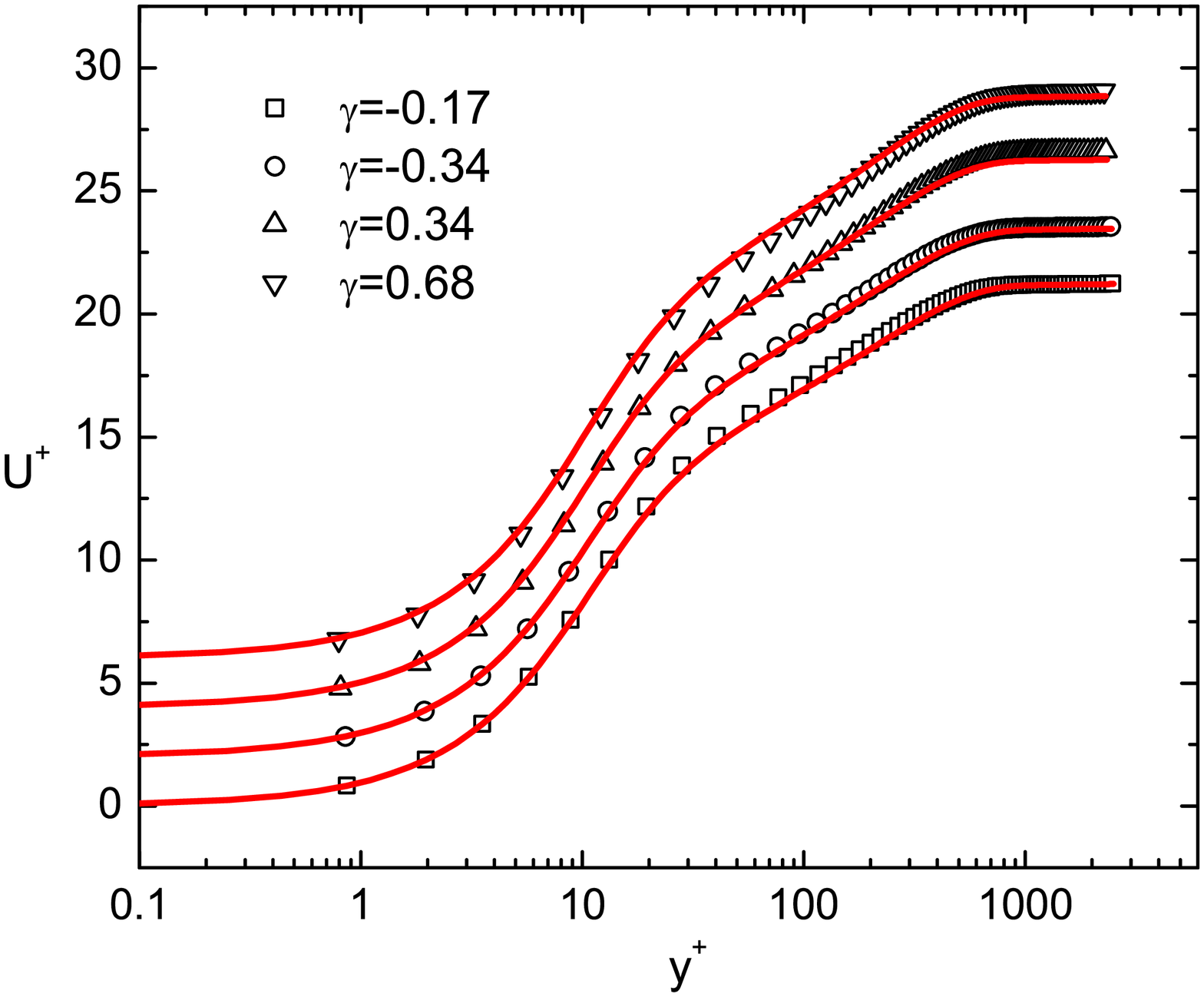}} 
\subfigure []{\includegraphics[width = 7 cm]{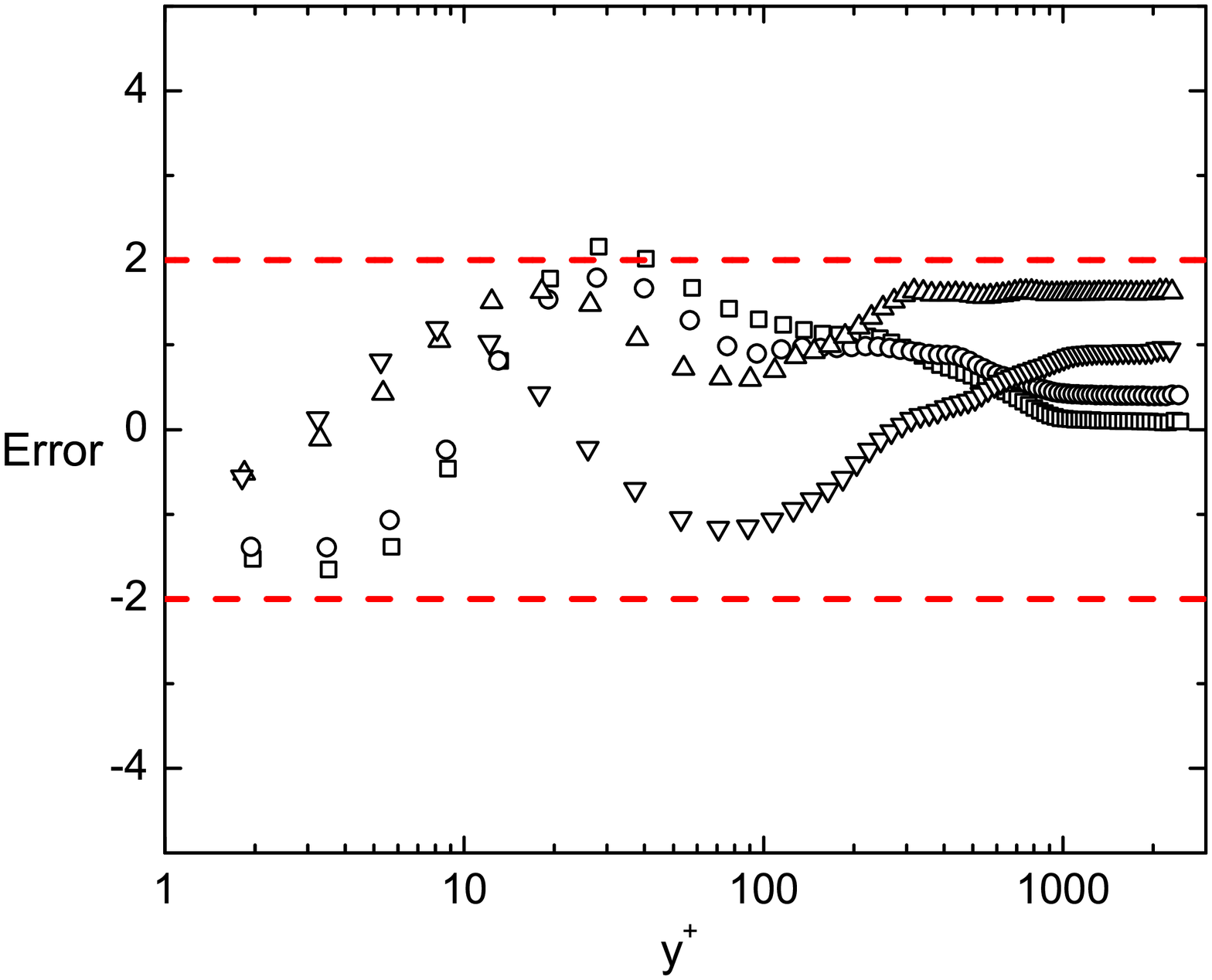}} 
  \caption{Verification of (\ref{eq:UT}). (a): Predicted $\hat {U}^+$ (lines) from $U^+_*$ by (\ref{eq:UT}) compared with DNS data (symbols). Each profile has been vertically shifted (by 2) for better display. (b): Relative errors (times 100) are uniformly bounded within 2\% for the entire flow region.}
  \label{fig:UT}
\end{figure}

The performance of the transformation (\ref{eq:transA}) is further illustrated by a reverse transformation from $U^+_*$ to $U^+$. In order words, we predict $U^+$'s at different $\gamma's$ from the single profile $U^+_*(y^+)$. Note that according to (\ref{eq:phiA}) and (\ref{eq:phiB}), we have
\begin{equation}\label{eq:phiC}
S_{}^ +  = {2}/[{{1 + \sqrt {1 + 4\xi^*(\xi^* - 1)({\ell^+/\ell^+_*})^2}} }]\approx{{2 }}/[ {1 +  \sqrt {1 + 4{}\xi^* (\xi^*  - 1){(1 + \eta\gamma )}^2} }]
\end{equation}
where $\xi^*=1/S^+_*$ (and $\eta=-1/9$). Thus, by integrating (\ref{eq:phiC}) with $y^+$, the resulted mean velocity is:
\begin{equation}\label{eq:UT}
\hat{U}^ {+}  = \int {S_{}^ +(\xi^*,\gamma) d{y^ + }}
\end{equation}
The results are shown in the figure \ref{fig:UT}. One can see the agreement is very good and the relative errors are within 2\% for the entire flow region. {Note that one may introduce a damping function} \citep{pope2000turbulent} to model $\ell^+$ and hence obtaining $U^+$. This is another topic to be presented elsewhere.

\begin{figure}
\subfigure []{\includegraphics[width = 7 cm]{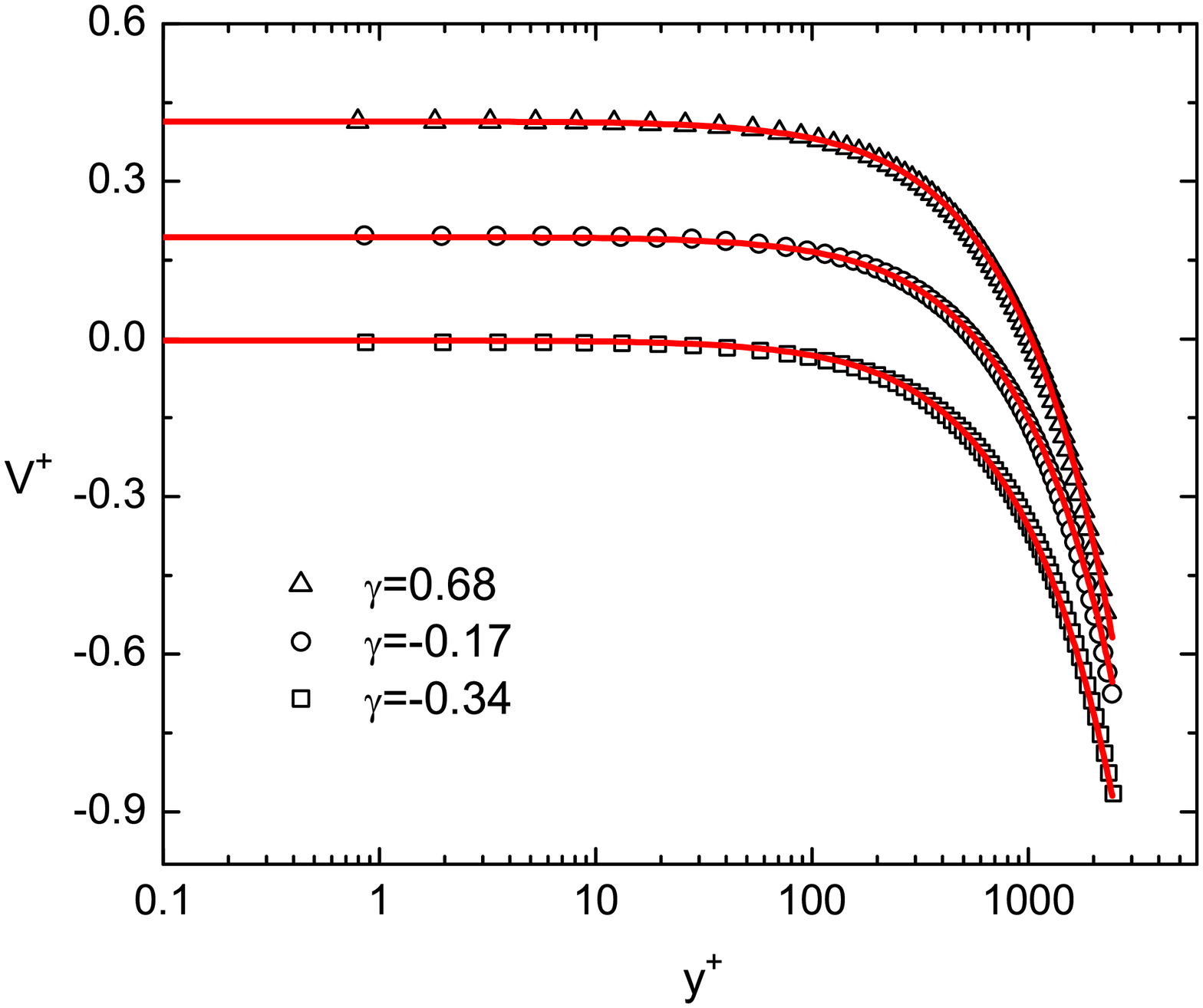}} 
\subfigure []{\includegraphics[width = 7 cm]{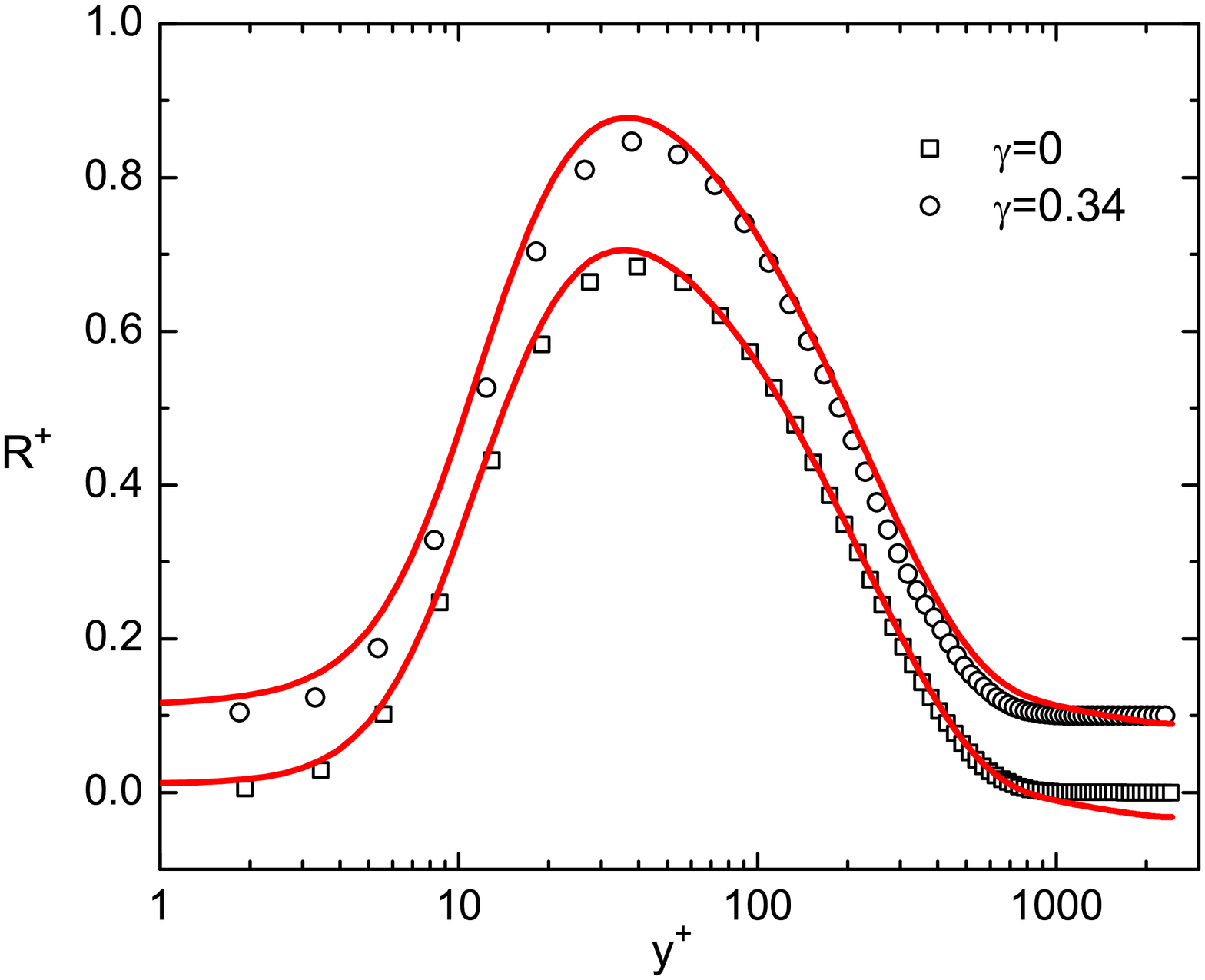}} 
  \caption{{Theoretical} $V^+$ from (\ref{eq:Vtrans}) shown in (a) (vertically shifted by 0.2 for each profile); and $R^+$ from (\ref{eq:Wtrans}) shown in (b) (vertically shifted by 0.1 for $\gamma=0.34$). Symbols are DNS data.}
  \label{fig:V}
\end{figure}

Moreover, the wall normal mean velocity and Reynolds shear stress are given based on the single $U^+_*{y^+}$ profile. According to (\ref{eq:UV}) we have
\begin{equation}\label{eq:Vtrans}
{\hat{V}^ + } = \gamma \sqrt {{K_p}} \hat{U}_{\infty }^ +  - {K_p}\hat{U}_{\infty }^ + {y^ + }\hat{U}^ +
\end{equation}
which indicates the wall normal velocity monotonically decreases from wall due to the sink flow constraint. In addition, the Reynolds shear stress from (\ref{eq:MMEutau}) is:
\begin{equation}\label{eq:Wtrans}
\hat {R }^ + =1- \partial_{y^+} \hat U^+-\hat {U}^{+3}_\infty K_p y^+ + \gamma \sqrt {{K_p}} \hat{U}_{\infty}^ + \hat{U}^+ + {K_p}\hat{U}_{\infty }^ + \int_0^{y^+} \hat{U}^{+2} dy'
\end{equation}
Above results are shown in the figure \ref{fig:V}, both $\hat{V}^+$ and $\hat{W}^+$ agreeing well with data. Note that towards freestream $\hat{R} ^+$ deviates modestly from data as it is very sensitive to the value of $K_p$; however, the agreement is remarkable on the whole. Note that from (\ref{eq:Vtrans}) and (\ref{eq:Wtrans}), the budget terms of (\ref{eq:MMEutau}) are calculated in figure \ref{fig:UV}a, in good agreement with data. Therefore, from $U^+_*$ we obtain a complete description of all the quantities in the equations (\ref{eq:mass}-\ref{eq:MME}). The results in turn support well the similarity transformation (\ref{eq:transA}).

\section{Conclusions and discussions}
We present a first similarity transformation for the mean velocities in sink flow TBL with blowing and suction effects. It achieves a mapping of different $U^+$'s at different blowing/suction strength $\gamma$'s into a universal $U^+_*$ at $\gamma=0$. The result builds on a Lie group symmetry analysis which derives the self-similarity equation (ODE) for the mean mass and momentum, and unifies the Falkner-Skan equation and the sink flow TBL with different pressure gradient and blowing/suction effects. {Unlike dimensional analysis where there are various ways to combine primary variables, the dilation symmetry here straightforwardly leads to the similarity variables. The latter has been emphasized by \cite{Cantwellbook} as an advantage of symmetry analysis and has been suggested to be used along with dimensional analysis.}

In the second part of this paper, a characteristic length is introduced whose first order expansion in $\gamma$ leads to an analytical expression of the transformation. {The latter is further shown to be a group invariant in the flow region where $S^+\approx1$ or $S^+\ll1$ under a generalized symmetry analysis}. Note that the expansion is key to the success of the transformation, which means that blowing/suction conditions can be effectively described by the relative variation (ratio) of length functions. Such a procedure (by characteristic lengths) has been introduced by \cite{shenjp} as a new way to quantify turbulent wall flows (with more results to be presented). Note that all of the mean flow quantities ($U^+$, $V^+$ and $R^+$) are calculated from the single profile $U^+_*(y^+)$ with high accuracy. The results indicate that the wall blowing/suction not only preserves the equilibrium condition but also leads to a new similarity among different $\gamma$'s.

A further discussion on the meaning of the dilation symmetry is presented here. Note that (\ref{eq:dilation}) means that if $U(x,y)$ is a solution of the sink flow TBL, $U^\ast=e^{-\varepsilon} U$, $x^\ast=e^{\varepsilon} x$, $y^\ast=e^{\varepsilon} y$ is also a solution. In other words, (\ref{eq:dilation}) implies a similarity solution of the form $U=x^{-1}f(y/x)$ for the sink flow TBL, just like the Blasius similarity solution $U=g(y/\sqrt{x})$ for the flat plate laminar boundary layer. As the former can be rewritten as $U^+=f(y^+)$ since $u_\tau\propto x^{-1}$ and $y^+=yu_\tau/\nu\propto y/x$, (\ref{eq:dilation}) also indicates that the mean velocity profiles at different streamwise locations would collapse  when nondimensionalized by $u_\tau$ and $y^+$ (applies also to the Reynolds shear stress profiles), which is actually confirmed by the DNS data. However, to emphasize, (\ref{eq:dilation}) does not result in any specific form of $U^+$ (nor of $R^+$). The latter of course allows the scaling such as the log law in the overlap region or the exponential law in the wake region as proposed by \cite{Oberlack2001}; nevertheless, these scalings are just `candidate' invariant solutions but not a direct consequence of the symmetry analysis. This is the difference between our current work and Oberlack's (2001).

This work also opens several important issues which are explained briefly. The first is on the $K_p$ effect. While the current study focuses on a specific $K_p$, more calculations are needed for different $K_p$'s to validate the transformation with an appropriate $\eta$. Secondly, towards a compete analytical description of mean velocity profiles (for instance the $U^+_*(y^+)$), {closure assumptions such as the mixing length model or the asymptotic logarithmic law can be introduced}. Along this direction, a third-parameter paradigm, i.e. $Re_\theta-K_p-\gamma$ is expected in analogy to \cite{jones2001}. Thirdly, it is important to delineate the $\gamma$ range where the similarity transformation holds, as a huge intensive blowing/suction effect would break the similarity transformation and the equilibrium flow state. This is already mentioned in the preceding expansion analysis where $|\gamma|<1$ is noted. Finally, a similar analysis can be carried out for the sink flow TBL with different roughness effects \citep{Yuan2014}. {Note that extending the current work to source flows would be interesting to determine if the rich variety of similarity flow states \citep{Moffatt1964,Moffatt1980}} would be broken by blowing/suction. All these to be pursued in future are essential to the fundamental understanding of turbulent wall flows.


\begin{acknowledgments}
This work benefitted from discussions with Z.S. She of Peking University. We thank O.N. Ramesh for sharing the DNS data from his PhD student S.S. \cite{Pat2014}.
\end{acknowledgments}


\appendix{\section{Symmetries of (\ref{eq:MMEpsi})}}
{More symmetries of (\ref{eq:MMEpsi}) can be calculated by using algebraical softwares. Below are the results by using Maple:
\begin{eqnarray}\label{eq:maple}
\begin{array}{l}
{\xi' _x} = {G_1}(x),\\
{\xi' _y} = (1 + \mathop {{G_1}}\limits^ \bullet  /2)y + {G_2}(x,{U_\infty },{V_w}),\\
{\eta' _\psi } = (\mathop {{G_1}}\limits^ \bullet  /2 - 1)\psi  + {G_3}(x),\\
{\eta' _R} =  - (\mathop {{G_1}}\limits^ \bullet  /2 + 3)R + {\partial _x}{G_4}(x,y) + {G_5}(x,{U_\infty },{V_w}),\\
{\eta' _{{U_\infty }}} =  - 2{U_\infty } + {\partial _y}{G'_4}/{U_\infty },\\
{\eta' _{{V_w}}} =  - (\mathop {{G_1}}\limits^ \bullet  /2 + 1){V_w} + \psi \mathop {{G_1}}\limits^{ \bullet  \bullet } /2 + \mathop {{G_3}}\limits^ \bullet.
\end{array}\end{eqnarray}
where $\xi'_i$ and $\eta'_i$ are infinitesimals for independent and dependent variables, respectively; the super script indicates $\mathop {{G}}\limits^ \bullet=dG/dx $ and $\mathop {{G}}\limits^ {\bullet\bullet}=d^2G/dx^2 $. Note that (\ref{eq:maple}) is equivalent to the dilation group (\ref{eq:dilation}) by letting $G_2=G_3=G_4=G_5=0$ and $G_1=-2x/\beta$. Also note that due to the boundary condition ($U=R=\psi=0$ at wall) with the fixed sink apex, no translation or rotation is permitted. In this paper, we focus on the dilation group (\ref{eq:dilation}) in analogy to the Blasius equation for laminar flows, and the resulting symmetry is sketched in figure \ref{fig:sink}. }

\end{document}